\documentclass{article}
\usepackage{spconf,amsmath,graphicx,hyperref}
\usepackage{algorithm}
\usepackage{algorithmic}
\usepackage[table,xcdraw]{xcolor}
\usepackage{booktabs}
\usepackage{pifont}

\copyrightnotice{U.S.\ Government work not protected by U.S.\ copyright}

\copyrightnotice{979-8-3503-2411-2/25/\$31.00 {\copyright}2025 Crown}

\copyrightnotice{979-8-3503-2411-2/25/\$31.00 {\copyright}2025 European Union}

\copyrightnotice{979-8-3503-2411-2/25/\$31.00 {\copyright}2025 IEEE}

\toappear{2026 IEEE International Workshop on Machine Learning for Signal Processing, Sep.\ 28-- Oct.\ 1, 2026, Atlanta, USA}

\title{From Who Said What to Who They Are: Modular Training-free Identity-Aware LLM Refinement of Speaker Diarization}
\name{Yu-Wen Chen$^{\dag}$$^1$, William Ho$^{\dag}$$^1$, Maxim Topaz$^2$, Julia Hirschberg$^1$, Zoran Kostic$^1$ \thanks{$^{\dag}$These authors contributed equally.}}
\address{
$^1$The Fu Foundation School of Engineering and Applied Science, Columbia University, United States\\
$^2$School of Nursing, Columbia University, United States
}

\begin{document}

\maketitle

\begin{abstract}
    Speaker diarization (SD) remains challenging in real-world scenarios due to dynamic environments and unknown speaker numbers. SD is rarely used alone and is typically paired with automatic speech recognition (ASR). However, existing non-modular SD+ASR frameworks lack flexibility and do not provide true speaker identities. We propose a training-free modular pipeline combining off-the-shelf SD, ASR, and a large language model (LLM) to determine who spoke, what was said, and who they are. Using structured LLM prompting on reconciled SD and ASR outputs, our method leverages semantic continuity in conversational context to refine low-confidence speaker labels and assigns role identities while correcting split speakers. On a real-world patient-clinician dataset, our approach achieves a 29.7\% relative error reduction over baseline reconciled SD and ASR. It enhances diarization performance without additional training and delivers a complete pipeline for SD, ASR, and speaker identity detection in practical applications.
\end{abstract}

\newcommand{\cem}[1]{\textcolor{blue}{cem: #1}}
\section{Introduction}
\label{sec:intro}
Speaker diarization (SD) identifies “who spoke when” in an audio recording~\cite{raghav2026diarizen}. While diarization models perform well in controlled environments, they face persistent challenges in real-world conditions~\cite{ryant2020third}. An SD system relying solely on acoustic cues is highly vulnerable to environmental variations, such as changes in a speaker’s voice when moving between rooms or interference from background noise such as a television. Also, most SD systems require specifying the maximum number of speakers, a setting highly sensitive in uncontrolled environments: over-estimation can split one speaker into distinct speakers, while under-estimation can merge different speakers into one. In practice, SD is rarely used alone. Most applications require not only knowing who spoke and when, but also what was said. Therefore, SD is commonly integrated with automatic speech recognition (ASR). Several studies have explored end-to-end joint training of ASR and SD models~\cite{shafey2019joint, park2020speaker, xia2022turn, cornell2024one}. Beyond a record of spoken words, ASR transcripts also capture semantic information that can supplement SD. More recently, advances in large language models (LLMs) have further enhanced diarization by leveraging their strong semantic understanding~\cite{paturi2024ag, kumar2025seal, yin2025speakerlm, cheng2026integrating}. 

Most prior methods employ end-to-end systems that integrate SD with ASR or LLMs. However, keeping these components as separate modules (i.e., a modular approach) offers several advantages, yet studies focusing on this design remain relatively limited. Modularization enables independent development and deployment, as ASR and SD systems are often trained on different datasets and developed by separate teams. It enables flexible, scalable integration of any ASR system providing word timings with diverse SD models. Moreover, joint modeling may degrade ASR, making modular architectures preferable for accurate transcription. Building on this insight, recent studies have proposed using a reconciliation or orchestration module to combine ASR and SD outputs, followed by LLM-based post-processing to address mismatches between the two. For example,~\cite{wang2024diarizationlm} leveraged LLMs to refine SD outputs, while~\cite{park2024enhancing} augmented an acoustic-based SD system with lexical information from an LLM. These studies demonstrate the potential of LLMs to enhance speaker diarization through post-processing. However, existing approaches involved fine-tuning LLMs for speaker diarization. This process requires additional training effort, and the resulting performance is highly influenced by the distribution of the fine-tuning data, raising concerns about the model's ability to generalize to out-of-domain scenarios. In contrast, leveraging general-purpose LLMs for speaker diarization without task-specific fine-tuning remains largely underexplored. Also, prior modular approaches do not consider identity detection, even though assigning generic labels such as \emph{spk0} or \emph{spk1}, rather than speaker identities like patient or clinician, is often insufficient in real-world scenarios.

In this study, we propose a training-free modular SD+ ASR+LLM pipeline that processes audio recordings to determine who spoke, when they spoke, what was said, and the identity of each speaker. Initial diarization and transcription alignments are generated using independent SD and ASR models, followed by leveraging a general LLM to improve diarization. The LLM suggests speaker labels for low-confidence segments using semantic context and infers speaker identities based on the entire conversation. This identity information, which is critical for downstream tasks, is found to be helpful for mitigating speaker-splitting errors that can significantly degrade SD performance in real-world settings, such as home healthcare recordings. Experimental results show that the proposed method improves diarization results with both open-source and commercial LLMs. The modular design of the proposed pipeline effectively leverages existing SD and ASR models from independent research without additional training, avoiding issues related to limited data availability and domain mismatch that can arise when SD, ASR, and role identification models are trained end-to-end. 


\section{Proposed method}\label{sec:method}

We select SD and ASR models that provide timestamps, allowing alignment between their outputs. We then introduce two approaches that directly leverage a general LLM to refine the initial SD+ASR results: (1) context-aware speaker label correction for addressing minor mis-assignments, and (2) identity detection for resolving major speaker-splitting errors. Details are presented below.

\subsection{Initial diarization and transcription}
We first run SD on the raw recording to obtain diarization segments and apply ASR to generate transcripts with word-level timestamps. To reconcile SD and ASR (denoted as SD+ASR), we match ASR words whose start times fall within the duration of each diarization segment. Mismatches arise in two cases: (1) an ASR word cannot be mapped to any diarization segment, in which case we create a new segment using the ASR timestamp and label it as an \emph{Unknown} speaker; (2) a diarization segment contains no ASR-recognized words, resulting in an empty transcript. We then re-run ASR on these mismatched segments. Segments labeled as \emph{Unknown} are retained only if the re-run ASR transcript is highly similar to the original (Levenshtein similarity $\geq$ 0.9), while empty segments are kept only if the ASR model recognizes at least one word in the reprocessing.

\subsection{Context-aware speaker label correction}

\begin{figure}[htbp!]  
\centering
\includegraphics[scale=0.88]{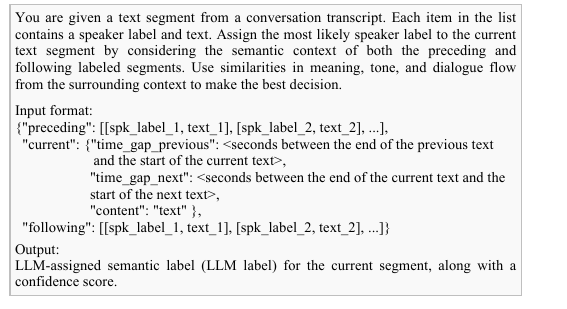}
\caption{Illustration of LLM prompt for context-aware speaker label correction.}
\label{fig:llm_prompt}
\end{figure}

Unlike previous studies that feed the entire conversation sequence into an LLM (e.g., {\ttfamily <spk1>} How are you doing {\ttfamily<spk2>} today? Good. \ldots), we employ a segment-level approach. At each step, the LLM is prompted to predict the speaker label of an SD segment (denoted as the LLM label). This segment-level approach avoids the difficulties with mapping whole-conversation-level outputs back to the original time alignment. We provide the LLM with each segment’s context and surrounding time gaps (Fig.~\ref{fig:llm_prompt}). Time-gap information is included because prior work has shown that LLMs can interpret temporal cues in text, such as assessing phrase breaks~\cite{wang2023assessing}. 

Considering the cost of LLM inference, we apply context-aware speaker label correction only to segments with low confidence. We compute each diarization segment’s speaker embedding and retrieve the top-k most similar segments in the same recording. Re-verified speaker labels are assigned based on the majority of speaker labels among these similar segments. Segments where the re-verified label differs from the original label are marked as low acoustic confidence. Words in low-acoustic-confidence segments, along with those labeled as ~\emph{Unknown} in the initial step, are further processed using an LLM.

\subsection{Identity detection}
Acoustic variations arising from environmental changes during a conversation can cause SD systems to erroneously split a single speaker into multiple pseudo-speaker labels. For example, shifts in a speaker’s position relative to the microphone can introduce noticeable acoustic variations, leading the SD system to incorrectly infer the presence of a new speaker. Such errors can substantially degrade diarization performance, as long contiguous segments are mis-attributed to different speakers. However, the semantic coherence of the discourse often suggests that the speaker’s identity remains consistent despite acoustic variability, offering valuable cues for resolving these errors.

We use an LLM to assign an identity to each speaker based on the full transcript and initial diarization labels. The prompt used is: \emph{``You are given a conversation with speaker labels. Your task is to assign an identity to each speaker. If different original speaker labels refer to the same person, merge them by assigning the same identity, but only do this if you are very sure.''} The LLM outputs a dictionary mapping each speaker label to an identity. For example, if both \emph{spk0} and \emph{spk1} are mapped to \emph{Patient}, it indicates that \emph{spk1} was erroneously split from the same speaker as \emph{spk0}. This allows us to detect cases where a single speaker has been mistakenly split into multiple labels and also provides the identity information needed for downstream tasks.

\subsection{Final aggregation}

We integrate the results of initial diarization, acoustic reverification, LLM speaker labeling, and identity mapping. The pseudo-code is shown in Algorithm~\ref{alg:llm_stage}. Finally, duplicated or nested segments are removed.

\begin{algorithm}
\caption{Final aggregation}
\begin{algorithmic}[1]
\FOR{each diarization\_segment}
    \IF{original\_label == \emph{"Unknown"}}
        \IF{LLM not confident \OR llm\_label == \emph{"Unknown"}}
            \STATE \textbf{continue} 
        \ELSE
            \STATE final\_label = map\_identity[llm\_label]
        \ENDIF
    \ELSIF{original\_label == reverified\_label \OR LLM not confident}
        \STATE final\_label = map\_identity[original\_label]
    \ELSE
        \STATE final\_label = MAJORITY\_VOTE(
        \STATE \quad \quad \quad \quad \quad map\_identity[original\_label], 
        \STATE \quad \quad \quad \quad \quad map\_identity[llm\_label], 
        \STATE \quad \quad \quad \quad \quad map\_identity[reverified\_label])
    \ENDIF
\ENDFOR
\end{algorithmic}
\label{alg:llm_stage}
\end{algorithm}

\section{Experimental setup}

\subsection{Data}

\textbf{Clinician–patient conversation data:} Under IRB approval, we collected clinician–patient conversations single-channel (48 kHz) audio recordings from real-world health home-care visits, which were then resampled to 16 kHz. We manually annotated ten samples (13.6–34.1 min, average 24.9 min) to produce ground-truth diarization segments. These recordings were selected to represent a variety of real-world challenges, including scenarios with high voice similarity between clinician and patient, unexpected speakers (e.g., family members), varying environments (e.g., changes in microphone position), and background noise (e.g., television). \\
\textbf{Meeting conversation data:}
We also evaluated on the public AMI-SDM (single distant microphone) version of the AMI Meeting Corpus~\cite{kraaij2005ami,carletta2005ami}, using the test split from~\cite{landini2022bayesian} (16 audios, 14.0–49.5 min, avg. 34.0 min). AMI was chosen because, like the clinician–patient data, identifying each speaker (e.g., which employee and their role) is essential, while also posing natural multi-party challenges such as overlapping speech, variable acoustics, and spontaneous turn-taking.

\subsection{Implementation details}
We used the speaker verification model \emph{TitaNet-Large}~\cite{koluguri2022titanet} to extract speaker embeddings and \emph{Parakeet-TDT-0.6B-v2}~\cite{parketeernvidia} for generating transcripts with word-level timestamps. For diarization, we evaluated two models: \emph{Sortformer-Diarizer-4spk-v1} (Sortformer)~\cite{parksortformer} and \emph{Pyannote Speaker-diarization-3.1}~\cite{Plaquet23}. Since Sortformer was trained on 90-second inputs, it does not generalize well to long recordings. To address this issue, we segmented each audio into shorter chunks, applied diarization independently, and then reconciled speaker labels across chunks. Specifically, we computed an average embedding for each predicted speaker within a chunk and clustered all average embeddings across chunks to align speaker labels. To determine the optimal chunk length, we performed a grid search on the AMI-SDM development split~\cite{landini2022bayesian}, where a 250-second chunk with a 5-second overlap yielded substantially lower DER than the 90-second setting. Since the number of speakers is unknown a priori, we set Sortformer’s maximum to its limit of four and used Pyannote’s default setting.

For speaker re-verification, we used Faiss~\cite{douze2024faiss} with the \emph{IndexFlatIP} setting and set the number of retrieved neighbors to 10. We tested our proposed method using three LLMs: \emph{GPT-4.1}~\cite{achiam2023gpt} (default settings), \emph{Qwen-3-8B}, and \emph{Qwen-3-32B}~\cite{yang2025qwen3}. The LLM’s confidence threshold is set to 0.9. In addition to the prompt described in Section~\ref{sec:method}, we added safeguards for \emph{Qwen-3} to enforce consistent output formatting, including using specific identities rather than general labels like “main” or “background,” and ensuring that all original speaker labels were considered. Also, for AMI-SDM, where multiple speaker may share similar roles (e.g., multiple developers), we ran \emph{Qwen-3} in thinking mode and expanded the identity detection prompt to require per-speaker evidence and more fine-grained role differentiation.
Evaluation followed standard SD procedures. We calculated the Diarization Error Rate (DER) using pyannote.metrics~\cite{pyannote.metrics} with a 0.25s collar. DER captures three types of errors: missed detection (ground-truth speech not detected), false alarms (speech detected but not in the ground-truth), and confusion (wrong speaker assigned). For the AMI corpus, with ground-truth transcripts available, we also report WER to evaluate transcription accuracy.

\section{Results}
\subsection{Performance on clinician-patient conversations}
\begin{table}[htbp!]
\centering
\caption{Performance on clinician-patient conversations. Abbreviations used in the table: \emph{“w trans.”} denotes with transcription, \emph{“FA”} denotes false alarm, \emph{“Conf.”} denotes confusion, and \emph{“Miss Det.”} denotes missed detection. All metrics are reported in percentages.}
\vspace{0.5em}
\label{tab:medical-general}
\renewcommand{\arraystretch}{1.2}
\resizebox{0.95\columnwidth}{!}{
\begin{tabular}{|c|c|c|c|c|c|}
\toprule \hline
 &
  w trans. &
  DER &
  FA &
  Conf. &
  Miss Det. \\ \hline \hline
\begin{tabular}[c]{@{}c@{}}SD\\[-0.4em] (Sortformer~\cite{parksortformer})\end{tabular}     & \ding{55} & 21.06          & 4.21          & 7.86          & 8.99          \\ \hline 
\begin{tabular}[c]{@{}c@{}}SD\\[-0.4em] (Pyannote~\cite{Plaquet23})\end{tabular}       & \ding{55} & 22.72          & 6.02          & 9.23          & 7.47          \\ \hline \hline \rowcolor[HTML]{EFEFEF} 
\begin{tabular}[c]{@{}c@{}}SD+ASR\\[-0.4em] (Sortformer)\end{tabular} & \ding{51} & 23.05          & 6.25          & 9.60          & \textbf{7.20} \\ \hline

\begin{tabular}[c]{@{}c@{}}SD+ASR+ \\[-0.4em] diarizationLM~\cite{wang2024diarizationlm}\end{tabular} & \ding{51} & 41.47          & 14.75          & 14.86          & 11.86 \\ \hline

\begin{tabular}[c]{@{}c@{}}\textbf{Proposed} \\[-0.4em] (Qwen-8B)\end{tabular}    & \ding{51} & 17.72          & 4.16          & 4.28          & 9.28 \\ \hline
\begin{tabular}[c]{@{}c@{}}\textbf{Proposed} \\[-0.4em] (Qwen-32B)\end{tabular}    & \ding{51} &  17.33       &   4.15       & 3.95          &    9.22   \\ \hline
\begin{tabular}[c]{@{}c@{}} \textbf{Proposed}\\[-0.4em] (GPT)\end{tabular}      & \ding{51} & \textbf{16.19} & \textbf{4.12} & \textbf{2.84} & 9.23                   \\ \hline \hline
  AWS &
  \ding{51} &
  29.29 &
  8.44 &
  11.34 &
  9.50 \\ \hline \bottomrule
\end{tabular}
}
\end{table}
Table~\ref{tab:medical-general} shows the performance on our real-world clinician-patient conversational data. We first evaluate raw diarization performance using two widely used open-source SDs: Sortformer and Pyannote. Since Sortformer outperforms Pyannote, we adopt it as the SD component in our framework. Combining SD and ASR (SD+ASR) increased diarization errors, as the systems are trained independently, leading to potential mismatches, but integrating diarization and transcription is essential, as we are interested in not only on who is speaking but also what is being said. 
Compared with the baseline SD+ASR, our method substantially reduces DER, even surpassing the performance of the original SD system. Also, it is effective with both commercial LLM (e.g., GPT) and open-source alternative (e.g., Qwen). In contrast, when applied to the SD+ASR outputs, diarizationLM~\cite{wang2024diarizationlm}, a previously proposed method, exhibited a notable decrease in performance. One primarily reason is that diarizationLM merges consecutive segments assigned to the same speaker. For example, the conversation “{\ttfamily <spk1>} hello {\ttfamily <spk1>} how are you?” may be merged into “{\ttfamily <spk1>} hello, how are you?”. In the original conversation, the separation between the two {\ttfamily<spk1>} segments reflects the long pause, but diarizationLM causes this timing information to be lost, which can negatively impact DER by including long silences. This also illustrates the time-alignment issues that arise when processing the entire conversation directly. Furthermore, diarizationLM preserves all pseudo-speaker labels and does not employ identity detection to address speaker mis-splits. As a final benchmark, we compare these results against a commercial SD+ASR service (AWS~\footnote{\url{https://aws.amazon.com/transcribe/}}) and our method achieves markedly better results.

\subsubsection{Ablation study of the proposed method}
Table~\ref{tab:ablation} presents our ablation study of the proposed method. First, re-verifying segments using labels from similar speaker embeddings yields results comparable to the original. If the original and re-verified labels mismatch, LLM-based labeling is used to support the final decision. Since GPT outperforms Qwen, we used it as the representative model for this ablation study. In GPT-full, segments take GPT’s label when the original and re-verified labels disagree. This increases confusion vs.\ the SD+ASR baseline, showing that diarization should not rely mainly on an LLM’s reasoning from dialogue context, which can be ambiguous for speaker identification. In contrast, GPT-refinement incorporates the GPT result only if it is highly confident, taking a majority vote over the original, re-verified, and GPT labels, which successfully reduces DER. GPT-identity merges speaker labels when GPT detects them as the same identity. This approach shows a significant improvement, highlighting the issue in SD models where the same speaker is mistakenly split into multiple roles under real-world conditions. Lastly, our proposed method, which combines GPT refinement with identity mapping and duplicated cleaning, achieves the best overall performance.

\begin{table}[htbp!]
\centering
\caption{Ablation study of the proposed method. Parentheses denote the previous step and \emph{“ref.”} is an abbreviation for refinement. All metrics are reported in percentages.}
\vspace{0.5em}
\label{tab:ablation}
\renewcommand{\arraystretch}{1.2}
\resizebox{0.95\columnwidth}{!}{
\begin{tabular}{|c|c|c|c|c|c|}
\toprule \hline
 &
  w trans. &
  DER &
  FA &
  Conf. &
  Miss Det. \\ \hline \hline
SD (Sortformer~\cite{parksortformer})                                                                   & \ding{55} & 21.06 & 4.21                      & 7.86 & 8.99          \\ \hline \hline
Re-verification                                                         & \ding{55} & 21.51 & 4.21                      & 8.30 & 8.99          \\ \hline \hline
\rowcolor[HTML]{EFEFEF} 
\begin{tabular}[c]{@{}c@{}}(SD+ASR)\\[-0.4em] Re-run ASR\end{tabular}            & \ding{51} & 21.63 & 4.54                      & 8.24 & 8.84 \\ \hline \hline
\begin{tabular}[c]{@{}c@{}}(Re-run ASR)\\[-0.4em] GPT-full\end{tabular} &
  \ding{51} &
    21.75 &
  4.50 &
  8.47 &
  8.77 \\ \hline
\begin{tabular}[c]{@{}c@{}}(Re-run ASR)\\[-0.4em] GPT-ref.\end{tabular}          & \ding{51} & 21.29 & 4.35                      & 7.93 & 9.01          \\ \hline
\begin{tabular}[c]{@{}c@{}}(Re-run ASR)\\[-0.4em] GPT-identity\end{tabular}      & \ding{51} & 16.61 & 4.59                      & 3.32 & 8.70          \\ \hline
\begin{tabular}[c]{@{}c@{}}(Re-run ASR)\\[-0.4em] GPT-ref.+identity\end{tabular} & \ding{51} & 16.42 & 4.35                      & 3.06 & 9.01          \\ \hline \hline
Proposed (GPT)                                                          & \ding{51} & 16.19 & 4.12 & 2.84 & 9.23          \\ \hline \bottomrule
\end{tabular}
}
\end{table}

\subsubsection{SD with LLM identity detection}
\begin{figure}[hpbt!]  
    \centering
    \includegraphics[scale=0.94]{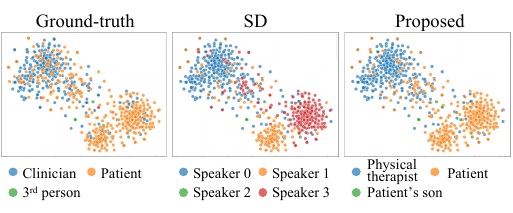}
    \caption{SD with LLM identity detection (t-SNE)}
    \label{fig:llm_identity}
\end{figure}
Fig.~\ref{fig:llm_identity}, visualizes speaker embeddings for each diarization segment of a single audio recording, with colors indicating distinct speakers. In SD results, three clear clusters can be observed, leading the SD model to assign each cluster a unique speaker label. However, according to ground-truth, SD labels for Speaker 1 and 3 actually correspond to the same speaker. Despite acoustic differences resulting from environmental changes, the conversation flow allows the LLM to recognize that speakers 1 and 3 are the same, mitigating SD errors from unknown speaker counts. Furthermore, the LLM assigns appropriate actual identity to the SD pseudo-labels, correctly identifying the clinician as a physical therapist, the patient, and the third speaker as the patient’s son, even without prior knowledge of the conversation.

Through manual inspection of the outputs, we observe that Qwen3-32B produces identity predictions that are much closer to those of GPT compared to Qwen3-8B. Specifically, GPT is more likely to provide both the speaker’s name (when available in the conversation) and their role, whereas Qwen3-8B provides less detailed identities and inconsistently uses names or roles, as illustrated in Table~\ref{tab:llm_identity_example}. All models improve diarization via identity detection, though GPT offers the largest benefits. In addition, GPT achieves strong performance with straightforward prompts, whereas Qwen models require specifically crafted instructions to reason and produce comparable identity accuracy.

\begin{table}[hpbt!]  
    \centering
    \caption{Example identity detection results across LLMs, with person names and organizations manually anonymized.}    
    \includegraphics[scale=1.]
    {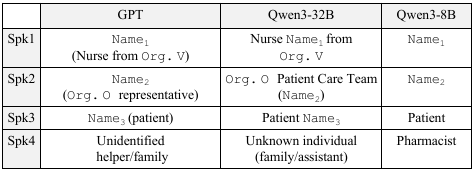}
    \label{tab:llm_identity_example}
\end{table}

\begin{table}[htbp!]
\centering
\caption{Performance on AMI-SDM reported in percentages.}
\vspace{0.5em}
\label{tab:meeting-general}
\renewcommand{\arraystretch}{1.2}
\resizebox{1.\columnwidth}{!}{
\begin{tabular}{|c|c|c|c|c|c|}
\toprule \hline
 &
  DER &
  FA &
  Conf. &
  Miss Det. &
  WER \\ \hline \hline
ASR             & -               & -               & -               & -               & 29.94 \\ \hline \hline
SD (Sortformer~\cite{parksortformer}) & 25.51          & 2.27         & 4.22         & 19.01          & -      \\ \hline
\cellcolor[HTML]{EFEFEF} \begin{tabular}[c]{@{}c@{}}SD+ASR\\ [-0.4em](Sortformer)\end{tabular} &
  \cellcolor[HTML]{EFEFEF}27.12 &
  \cellcolor[HTML]{EFEFEF} 3.91 &
  \cellcolor[HTML]{EFEFEF} 5.94 &
  \cellcolor[HTML]{EFEFEF} 17.27 &
  \cellcolor[HTML]{EFEFEF} 32.54 \\ \hline
  Proposed (Qwen3-8B)   & 25.51  & 1.84 & 4.82 & 18.86 & 31.70 \\ \hline
  Proposed (Qwen3-32B)  & 24.55  & 2.00 & 4.23 & 18.32 & 32.32 \\ \hline
  Proposed (GPT)        & 24.51  & 2.01 & 4.27 & 18.23 & 32.31 \\ \hline 
  diarizationLM~\cite{wang2024diarizationlm}  & 38.67  & 7.00 & 14.93 & 16.74 & 33.36 \\ \hline \hline
SD (Pyannote~\cite{Plaquet23})   & 18.26 & 2.56          & 6.24          & 9.46 & -      \\ \hline
 \cellcolor[HTML]{EFEFEF} \begin{tabular}[c]{@{}c@{}}SD+ASR\\ [-0.4em](Pyannote)\end{tabular} &
  \cellcolor[HTML]{EFEFEF} 18.92 &
  \cellcolor[HTML]{EFEFEF} 3.23 &
  \cellcolor[HTML]{EFEFEF} 6.51 &
  \cellcolor[HTML]{EFEFEF} 9.18 &
  \cellcolor[HTML]{EFEFEF} 32.16 \\ \hline 
  Proposed (Qwen3-8B)   & 20.69  & 2.38 & 7.52 & 10.79 & 31.48 \\ \hline
  Proposed (Qwen3-32B)  & 19.48  & 2.44 & 6.82 & 10.21 & 31.76 \\ \hline
  Proposed (GPT)        & 18.37  & 2.49 & 5.99 & 9.90  & 31.92 \\ \hline
  diarizationLM         & 74.18  & 5.99 & 37.89 & 30.30 & 47.49 \\ 
\hline\bottomrule
\end{tabular}
}
\end{table}

\subsection{Performance on meeting conversations}
Table~\ref{tab:meeting-general} shows performance on meeting conversations. Again, mismatches between SD and ASR increase DER for SD+ASR compared to SD alone. While our proposed refinement shows improvement when applied to Sortformer, Pyannote's DER degrades slightly when combined with Qwen. Unlike clinician–patient data, meetings contain more homogeneous speaker identities (e.g., all “developers"), causing smaller LLMs (e.g., Qwen-8B \& 32B) to struggle in distinguishing individual speakers and leading to errors when merging segments by identity. Nevertheless, for both SD systems, our proposed (GPT) consistently reduces DER compared to the SD+ASR baselines, demonstrating its effectiveness as a generalizable post-processing approach for diverse SD models and across datasets.

\section{Conclusion}
We propose a training-free modular pipeline that integrates LLM-based speaker identity detection into SD and ASR. Important for real-world applications, our identity detection requires no prior knowledge and helps mitigate a key challenge in SD systems: the unknown number of speakers. Together with other components, the design enables robust diarization improvements in dynamic environments. On real-world clinician–patient data, our method achieved a 29.7\% relative DER reduction over an SD+ASR baseline, with consistent gains across datasets, SD models, and LLMs. In contrast to widely studied end-to-end systems, the modular architectures offer a practical alternative, allowing flexible substitution and robust adaptation to new domains without training.

\section{Acknowledgment}
This work was supported by the National Institute of Health under Grant No. 1R01AG081928-01. 


\bibliographystyle{IEEEbib}
\bibliography{strings,refs}

\end{document}